\documentclass[aip, apl,amsmath,amssymb,reprint, twocolumn]{revtex4}
\usepackage{graphicx}
\begin{document}
\title{Counting Near Infrared Photons with Microwave Kinetic Inductance Detectors}

\author{W. Guo$^{1}$}
\author{X. Liu$^1$}
\author{Y. Wang$^{1,2}$\footnote{Electronic mail: qubit@home.swjtu.edu.cn}}
\author{Q. Wei$^1$}
\author{L. F. Wei$^{1,3}$\footnote{Electronic mail: weilianfu@gmail.com}}
\author{J. Hubmayr$^2$}
\author{J. Fowler$^2$}
\author{J. Ullom$^2$}
\author{L. Vale$^2$}
\author{M. R. Vissers$^2$}
\author{J. Gao$^{2}$}

\affiliation{
1) Quantum Optoelectronics Laboratory, School of Physical Science and Technology, Southwest Jiaotong University, Chengdu, 610031, China\\
2) National Institute of Standards and Technology, Boulder, CO 80305, USA\footnote{Contribution of the U.S. government, not subject to copyright}\\
3) State Key Laboratory of Optoelectronic Materials and Technologies, School of Physics, Sun Yat-Sen University, Guangzhou 510275, China}
\date{\today}

%\author{W. Guo}
%\affiliation{Quantum Optoelectronics Laboratory, Southwest Jiaotong University, Chengdu, Sichuan 610031, China}
%\author{X. Liu}
%\affiliation{Quantum Optoelectronics Laboratory, Southwest Jiaotong University, Chengdu, Sichuan 610031, China}
%\author{Y. Wang}
%\altaffiliation{Correpondence E-mail: qubit@home.swjtu.edu.cn}
%\affiliation{Quantum Optoelectronics Laboratory, Southwest Jiaotong University, Chengdu, Sichuan 610031, China}
%\affiliation{National Institute of Standards and Technology, Boulder, CO 80305, USA}
%\author{Q. Wei}
%\affiliation{Quantum Optoelectronics Laboratory, Southwest Jiaotong University, Chengdu, Sichuan 610031, China}
%\author{L. F. Wei}
%\altaffiliation{Correpondence E-mail: weilianfu@gmail.com}
%\affiliation{Quantum Optoelectronics Laboratory, Southwest Jiaotong University, Chengdu, Sichuan 610031, China}
%\affiliation{State Key Laboratory of Optoelectronic Materials and Technologies, School of Physics, Sun Yat-Sen University, Guangzhou 510275, China}
%\author{J. Hubmayr}
%\affiliation{National Institute of Standards and Technology, Boulder, CO 80305, USA}
%\author{L. Vale}
%\affiliation{National Institute of Standards and Technology, Boulder, CO 80305, USA}
%\author{M. R. Vissers}
%\affiliation{National Institute of Standards and Technology, Boulder, CO 80305, USA}
%\author{J. Gao}
%\affiliation{National Institute of Standards and Technology, Boulder, CO 80305, USA}
%\date{\today}

\begin{abstract}
We demonstrate photon counting at 1550~nm wavelength using microwave kinetic inductance detectors (MKIDs) made from TiN/Ti/TiN trilayer films with superconducting transition temperature $T_{c} \approx$ 1.4~K. The detectors have a lumped-element design with a large interdigitated capacitor covered by aluminum and inductive photon absorbers whose volume ranges from 0.4~$\mu$m$^3$ to 20~$\mu$m$^3$. The energy resolution improves as the absorber volume is reduced. We achieved an energy resolution of 0.22~eV and resolved up to 7 photons per optical pulse, both greatly improved from previously reported results at 1550~nm wavelength using MKIDs. Further improvements are possible by optimizing the optical coupling to maximize photon absorption into the inductive absorber.
\end{abstract}

\maketitle

%The photon-number-resolving (PNR) detectors show monotonic response (usually linear in the operating range) to the photon number (energy), therefore they are able to count the number and measure the energy of individual photons. Particularly

Photon-number-resolving (PNR) detectors at near infrared wavelengths have important applications in a number of frontier fields, such as quantum secure communications~\cite{Hiskett}, linear optical quantum computing~\cite{Knill} and optical quantum metrology~\cite{Zwinkels}. Compared to more conventional detectors at this wavelength, such as silicon-based detectors~\cite{Finger}, superconducting detectors have lower dark-count rate, higher sensitivity, and broadband response. They show great promise in serving as the basic building blocks for efficient PNR devices. For example, by spatial or temporal multiplexing of superconducting nanowire single-photon detectors (SNSPDs)~\cite{Tsman,Divochiy,Dauler,Mattioli}, photons can be counted at high speed. But the single-element nanowire has no intrinsic PNR and energy-resolving capabilities. Alternatively, single-element transition edge sensors (TESs)~\cite{Irwin} have demonstrated high quantum efficiency and multi-photon discrimination at telecommunication wavelengths~\cite{Miller,Lita,Calkin}. Recently, counting up to 29 photons and intrinsic energy resolution $\approx 0.11$~eV at $1550$ nm wavelength have been achieved in Ti/Au TESs~\cite{Lolli,Lolli2,Lolli3}.

Another type of superconducting detector possessing intrinsic photon-number-resolving and energy-resolving power is the microwave kinetic inductance detector (MKID)~\cite{Day}. MKIDs are cooper pair breaking detectors based on high-quality factor (high-$Q$) superconducting resonators~\cite{zmuidzinas,yiwen}. The absorption of a photon with energy higher than twice the gap energy ($h\nu>2\Delta$) can break Cooper pairs into quasiparticles, changing the surface impedance of the resonator and resulting in a lower resonance frequency $f_{r}$ and higher internal dissipation (or lower quality factor $Q_{i}$). When applying a short optical pulse to the detector and probing the resonator with a microwave tone near the resonance frequency, one can obtain a pulse response in the  complex forward transmission $S_{21}$, as shown in Fig. 1(a). This photon response can be measured using a homodyne detection scheme (Fig.~1(d)) and the signal can be decomposed into frequency and dissipation responses (Fig.~1(a),(b)) for pulse analysis.

\begin{figure}[ht]
\includegraphics[width=8.5cm]{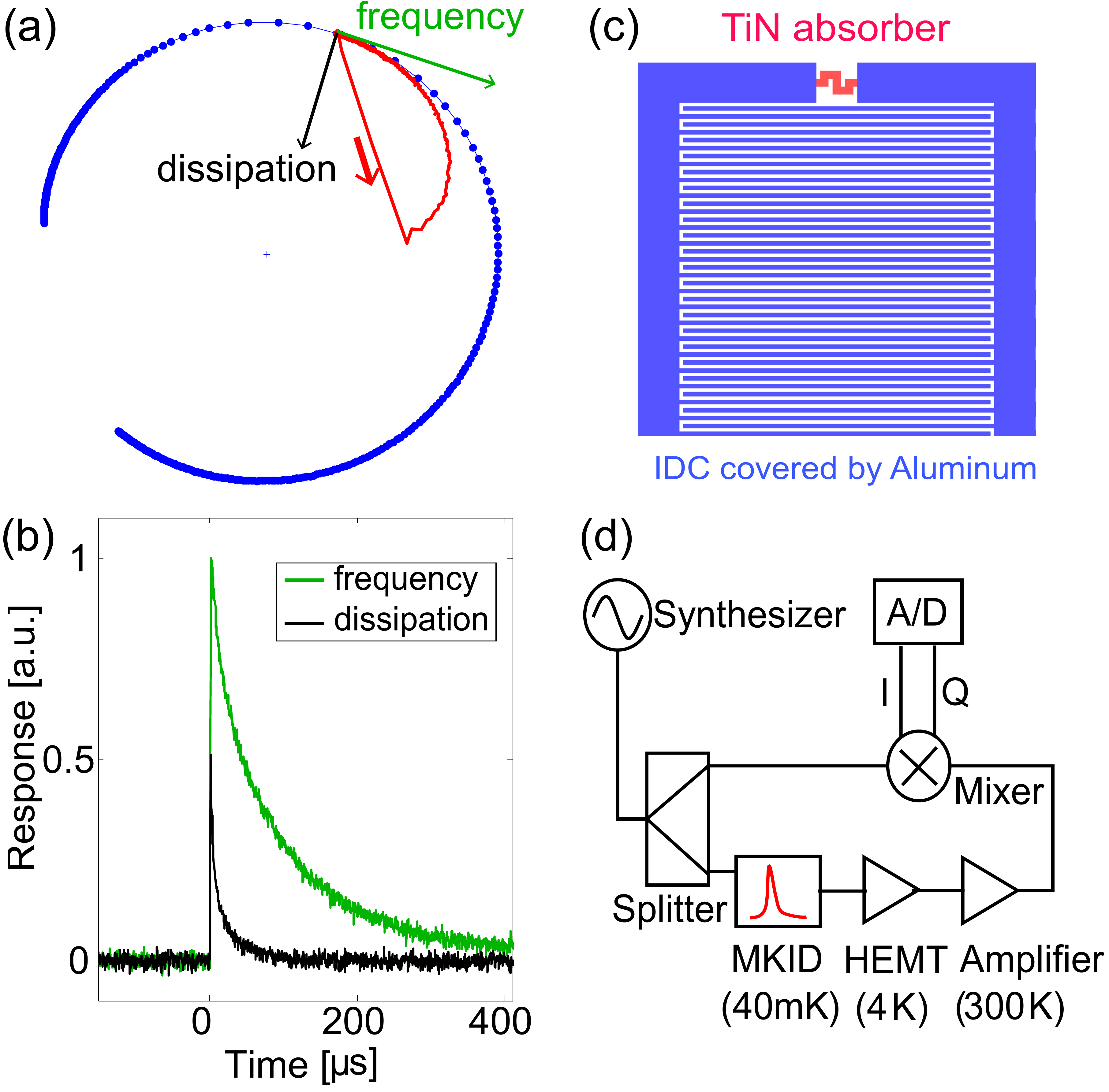}
\caption{[Color online] (a) Pulse response in the complex $S_{21}$ plane. The blue circle represents the resonance loop from a frequency sweep. The red line is the averaged response pulse after photon absorption and the red arrow shows the rising-edge of the pulse. This response can be projected to frequency and dissipation responses, with directions tangent and normal to the resonance loop. (b) The averaged frequency and dissipation pulse responses in time domain. (c) A schematic of the MKID design. The resonator has a lumped-element design, with a small volume of meandered inductive strip (red) in parallel with a large interdigitated capacitor (IDC), which is capped by a layer of aluminum (blue). (d) Homodyne detection scheme used to read out MKIDs.}
\label{fig:cavity}
\end{figure}

 Compared to TESs, MKIDs are easy to fabricate and multiplex into large arrays. A large array of MKIDs can be measured using a pair of coaxial cables, which greatly reduces the complexity of the instrument design. Previously, MKIDs with PNR capability have mostly been considered for astronomy applications at the visible wavelength~\cite{Bumble}. Single-photon counting at telecommunication wavelengths (near infrared) with titanium-nitride (TiN) MKIDs was first demonstrated in Ref.~\cite{Jiansong}, where a full-width-at-half-maximum (FWHM) energy resolution $\Delta E \approx$ 0.4~eV was achieved and up to 2-photon events were resolved. In this letter, we present an optimized MKID design based on TiN/Ti/TiN trilayer films and improved photon counting performance at 1550~nm wavelength: energy resolution $\Delta E \approx$ 0.22~eV is obtained and up to 7-photon events can be resolved.

 %We also systematically study the detector performance by varying the absorber volume $V$. We find that the device responsivity in fractional frequency shift scales linearly with $1/V$, resulting that both the energy-resolving power and the photon number that can be resolved decrease with increasing inductor volume $V$. Our results may provide a feasible approach to further improve the photon-counting with MKIDs.

Our detectors are made from a 20~nm thick TiN/Ti/TiN trilayer film~\cite{Vissers} ($T_{c} \approx 1.4$~K) deposited on a high-resistivity Si substrate. Such TiN trilayer films were initially developed for feedhorn-coupled MKIDs which have recently demonstrated photon-noise limited sensitivity at submillimeter wavelengths~\cite{hannes}. As shown in Fig.~1(c), our detectors comprise a large IDC shunted by a meandered inductive strip.  The latter serves as a sensitive photon absorber. The IDC area is $\approx$ 0.7~mm $\times$ 0.7~mm, with $5$ $\mu$m finger/gap width. This large area IDC is used to suppress the two-level system (TLS) noise in the substrate~\cite{gao2008}. The IDC is covered with a 100~nm-thick layer of aluminum (Al). Because of the low current density in the IDC and the much lower kinetic inductance of Al than TiN, the response from a photon hitting the IDC area is negligible. We designed $13$ resonators on a 10~mm by 5~mm chip, with inductor strip width ranging from 1~$\mu$m to 20~$\mu$m, length from 10~$\mu$m to 100~$\mu$m and volume from 0.4~$\mu$m$^3$ to 20~$\mu$m$^3$, to systematically study the dependence of the detector performance on the absorber geometry. All the resonance frequencies are designed to be around $6$~GHz and all the resonators are coupled to a common microstrip feedline with coupling quality factor $Q_{c} \approx 1.5 \times 10^{4}$.

%We also minimize the volume of the inductive strip where the photons are expected to be absorbed in order to increase responsivity.

%\begin{table}[ht]
%\caption{Chip-A design.}
%\label{tab:Multimedia-Specifications}
%\begin{center}
%\begin{tabular}{|l|l|l|l|l|l|l|l|l|l|l|}
%\hline
%\rule[-1ex]{0pt}{0.5ex}  Resonator & 1 & 2 & 3& 4 & 5 & 6& 7 & 8 & 9 \\
%\hline
%\rule[-1ex]{0pt}{1.5ex}  Width & 10.8 & 10.8 & 10.8& 10.8 & 10.8 & 10.8& 10.8 & 10.8 & 10.8 \\
%\hline
%\rule[-1ex]{0pt}{1.5ex}  Sqs & 10.8 & 10.8 & 10.8& 10.8 & 10.8 & 10.8& 10.8 & 10.8 & 10.8 \\
%\hline
%\rule[-1ex]{0pt}{1.5ex}  Volume & 10.8 & 10.8 & 10.8& 10.8 & 10.8 & 10.8& 10.8 & 10.8 & 10.8 \\
%\hline
%\end{tabular}
%\end{center}
%\end{table}

The detectors are cooled in a dilution refrigerator to a base temperature of 40~mK. At this temperature, the internal quality factors of the resonators are measured to be around $10^5$. A 1550~nm laser diode driven by a function generator at room temperature is used to generate optical pulses with a width of 200~ns at a repetition frequency of 120~Hz. The incident photons are then attenuated and guided into the device box mounted at the mixing chamber stage through a bare optical fiber. In this demonstration experiment, we did not optimize the optical coupling to the absorber and the light exiting the fiber flood illuminates the entire chip instead of being focused only onto the absorber area. As a result, the optical efficiency is rather low, which we plan to improve in future experiments. As shown in Fig.~1(d), the standard homodyne scheme is used to read out the resonators. We probe the resonators at a microwave frequency that maximizes the frequency response $\delta S_{21}/\delta f_r$ and the microwave power is chosen to be 2~dB below bifurcation power to avoid the strong non-linear effects~\cite{zmuidzinas} in the resonator. For each optical pulse, the corresponding response of the detector is digitized at a sampling rate of 2.5~Ms/s. The raw data are converted to the frequency and dissipation responses. Only the frequency response data are further analyzed, because the dissipation response is smaller compared to the frequency response and the dissipation pulse decay time is much faster (see Fig1.~(b))due to the anomalous electrodynamic effect found previously in TiN films~\cite{PropTiN, Jiansong,hannes}. Note that we have used a rigorous non-linear fitting procedure to directly convert the pulse trajectory in the IQ plane to the fractional frequency shift, because the response in fractional frequency shift unit is always linearly proportional to the change in the quasiparticle density, even when the pulse response is large (approaching the resonator line-width) and the phase shift becomes nonlinear. We analyze the pulse data by using standard Weiner optimal filter procedures and the filtered pulse height data are used to generate photon-counting statistics.

%with a high electron mobility transistor (HEMT) amplifier (noise temperature $T_{n} \sim$  $3$ K).
%to ensure the photon arrival rate smaller than the device bandwidth so we can count photons. Then describe a little data analysis based on nonlinear fitting......
\begin{figure}[ht]
\includegraphics[width=8.5cm]{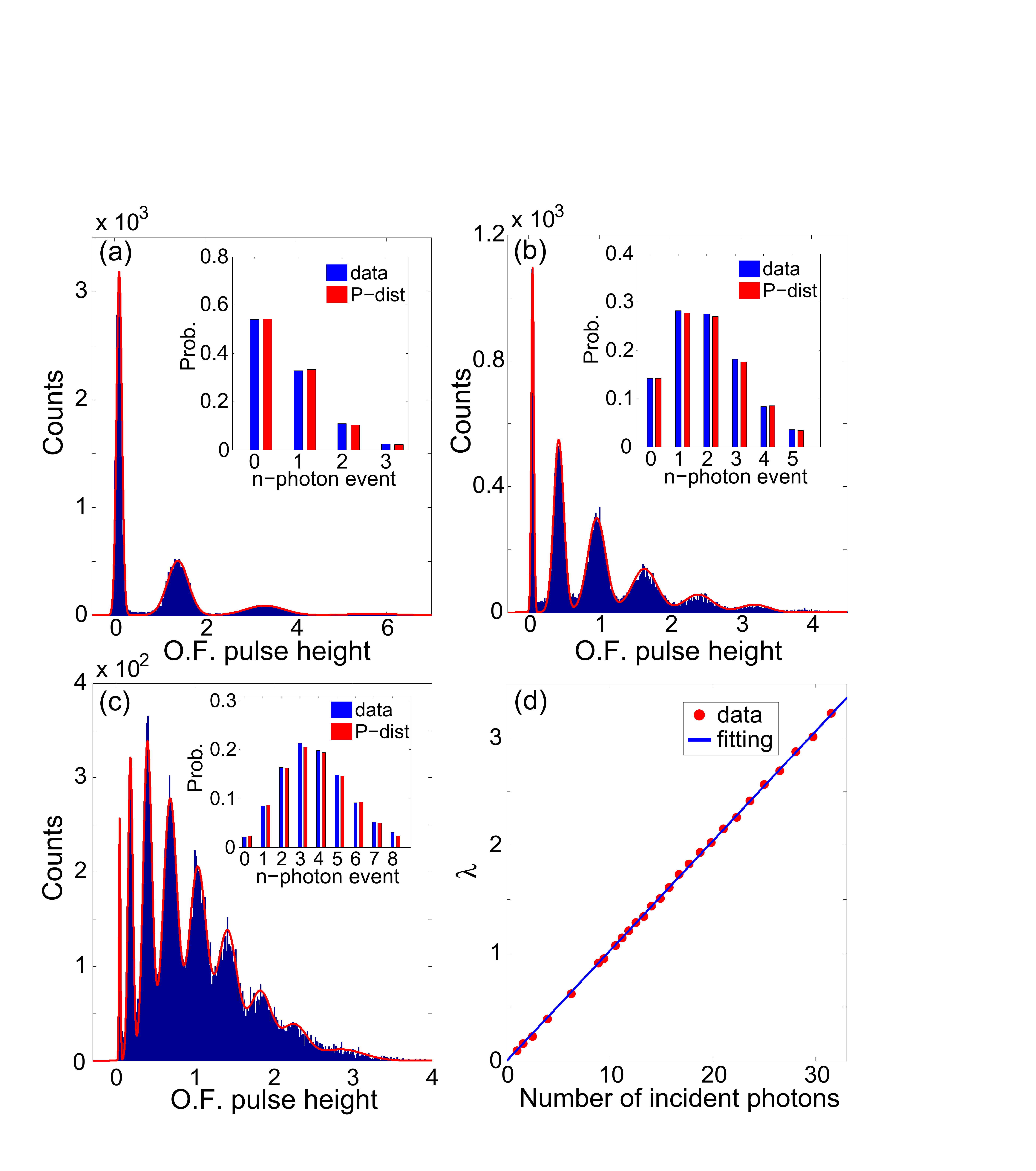}
\caption{[Color online] (a) A histogram of the optimally filtered (O.F.) pulse height (normalized by the template pulse) using frequency readout. A 4-peak Gaussian fit to the
data is shown by the red line. Inset: the probability of the $n$-photon event (calculated by the area in each Gaussian peak normalized
by the total area) fit to a Poisson distribution with $\lambda =$ 0.61. (b) Photon counting histogram ($\lambda =$ 1.95), fit by a superposition of 6 Gaussian peaks. (c) Photon counting histogram ($\lambda =$ 3.78) where 7-photon events are resolved. (d) The detected mean photon number per pulse (red dots) vs. the estimated total number of incident photons onto the absorber area. The slope of the linear fitting (blue curve) suggests the photon-device coupling efficiency is $\approx 10 \%$. }
\label{fig:cavity}
\end{figure}

Fig.~2(a) shows a histogram of the optimally filtered pulse height data for 2$\times 10^4$ pulse events measured from the resonator with absorber width of 2~$\mu$m and volume of 1.92~$\mu$m$^3$. The first 3 peaks, which correspond to the events of 0, 1, and 2 photons being absorbed in the detector, are clearly observed. We fit the histogram to a model of a superposition of 4 Gaussian peaks with independent heights and widths, as shown by the red profile in Fig.~2(a). The FWHM energy resolution $\Delta E_{n}$ of the $n$-photon peak is related to the standard deviation $\sigma_{n}$ of the $n$-th Gaussian peak by:
\begin{equation}
\Delta E_{n}=2\sqrt{2ln(2)}\frac{\sigma_{n}}{A_{n}-A_{n-1}}h\nu,~n=1,2,...
\end{equation}
where $h\nu = 0.80$~eV is the energy of a single 1550~nm photon and $A_{n}$ is the pulse height of the $n$-photon peak. The obtained FWHM energy resolutions for the 1-photon and 2-photon peaks are %$\Delta E_{0} =$ 0.088~eV,
$\Delta E_{1} =$ 0.34~eV and $\Delta E_{2} =$ 0.42~eV respectively. Here we claim a peak is resolved if $\Delta E/h\nu<1$. According to this criterion, this detector has the sensitivity to resolve the first 3 peaks (0-, 1- and 2-photon). According to the stochastic nature of the photon detection process, the $n$-photon events should obey Poisson statistics. Indeed, as shown in the inset of Fig.~2(a), the counts in the $n$-photon peak (proportional to the area of each Gaussian) normalized by the total counts match a Poisson distribution with $\lambda =$ 0.61. $\lambda$ is the mean photon number absorbed by the detector, suggesting that our detector detects an average of 0.61 photons per pulse event.

Fig.~2(b) shows the photon counting histogram at a higher input optical power, corresponding to a mean photon number $\lambda =$ 1.95. The first 6 (0- to 5-photon) peaks are resolved with the energy resolutions of
%$\Delta E_{0} =$ 0.092~eV,
$\Delta E_{1} =$ 0.36~eV and $\Delta E_{2} =$ 0.45~eV for the 1-, and 2-photon peak respectively, which both slightly increase from Fig.~2(a). Fig.~2(c) shows the histogram at an even higher optical power with a mean photon number of $\lambda =$ 3.78, where the first 8 (0- to 7-photon) peaks are resolved.
%The increasing overlap between photon peaks with higher mean photon number indicates that the maximum number of resolved photons is affected by the degradation of the energy resolution $\Delta E_{n}$ with increasing $n$ and input optical power.

In the 3 histograms shown in Fig.~2, we see that the 1- and 2-photon peaks are clearly broadened as compared to the 0-photon peak, indicating that additional noise arises when photons are absorbed and the energy resolution for the $n$-photon peak ($n \geq 1$) is not dominated by the background noise of the detector in the dark environment. We speculate the broadening might be related to several factors, including position-dependent response of the absorber, parasitic response from the photons hitting the non-absorber area (e.g., IDC, substrate, feedline), some unknown sources of photon-induced noise. We have simulated the current distribution using Sonnet (an electromagnetic simulation software) and the results show that the current is very uniform throughout the inductor strip to be within 0.4$\%$. This is expected because the dimensions of the inductors ($<100~\mu$m) are much smaller than the microwave wavelength ($>1$ cm around 6 GHz). Since the resonator frequency response is proportional to the local kinetic inductance change weighted by the square of the current distribution~\cite{gaothesis}, broadening of the photon peak due should not be dominated by the non-uniform current distribution in the inductive absorber.
%To study the position-dependent response from the absorber, we simulated the current distribution and found that the sectional current is very uniform along the inductor. The difference between the current at the inductor center and the current at the edge is below 0.4$\%$. This is reasonable because the longest inductor is only 100~$\mu$m, far smaller than the 6~GHz microwave wavelength (>2~cm). Since the response is proportional to the local kinetic inductance change and weighted by the square of the current distribution~\cite{gaothesis}, we think the position dependent response from the inductor gives negligible contribution to the broadening of the photon peak width because of the good current uniformity.

%This broadening might be partly attributed to the photon response from the parts other than absorber, as photons are hitting all areas (e.g., IDC, Si substrate, feed-line, etc.) in the device without restriction. This may also explain the small positive shift of the 0-photon peak from origin. However, this shift is too small compared to the 1-photon peak width, suggesting the non-absorber response is not enough to fully account for this broadening. We speculate that photon-induced noise or position dependent response from absorber are present in our device, which need to be further investigated.

% Second, the localized variations in the superconducting energy gap across the absorber can also cause a position dependent photon response.
In Fig.~2(d), we plot the detected mean photon number as a function of the estimated total number of photons incident onto the absorber area, which is perfectly linear as expected. The incident photon number is estimated from the total optical power measured by a power meter and the solid angle covered by the absorber area at the distance from the absorber to the fiber tip. Due to the low photon absorption efficiency, our detector can absorb and detect only 1 photon for approximately $10$ incoming photons hitting the absorber.

In this work, we have 13 resonators with different absorber volumes, which allows us to compare the photon counting statistics. The main results are summarized in Fig.~3. Fig.~3(a) shows the 1-photon responsivity (fractional frequency shift $\delta f_{r}/f_{r}$ induced by absorbing 1 photon) as a function of the absorber volume $V$. The measured responsivity is fitted well by a linear relation with $1/V$. This is expected because $\delta f_{r}/f_{r} \propto \delta n_{qp} \propto 1/V$, where $n_{qp}$ is the quasiparticle density.

\begin{figure}[ht]
\includegraphics[width=8.5cm]{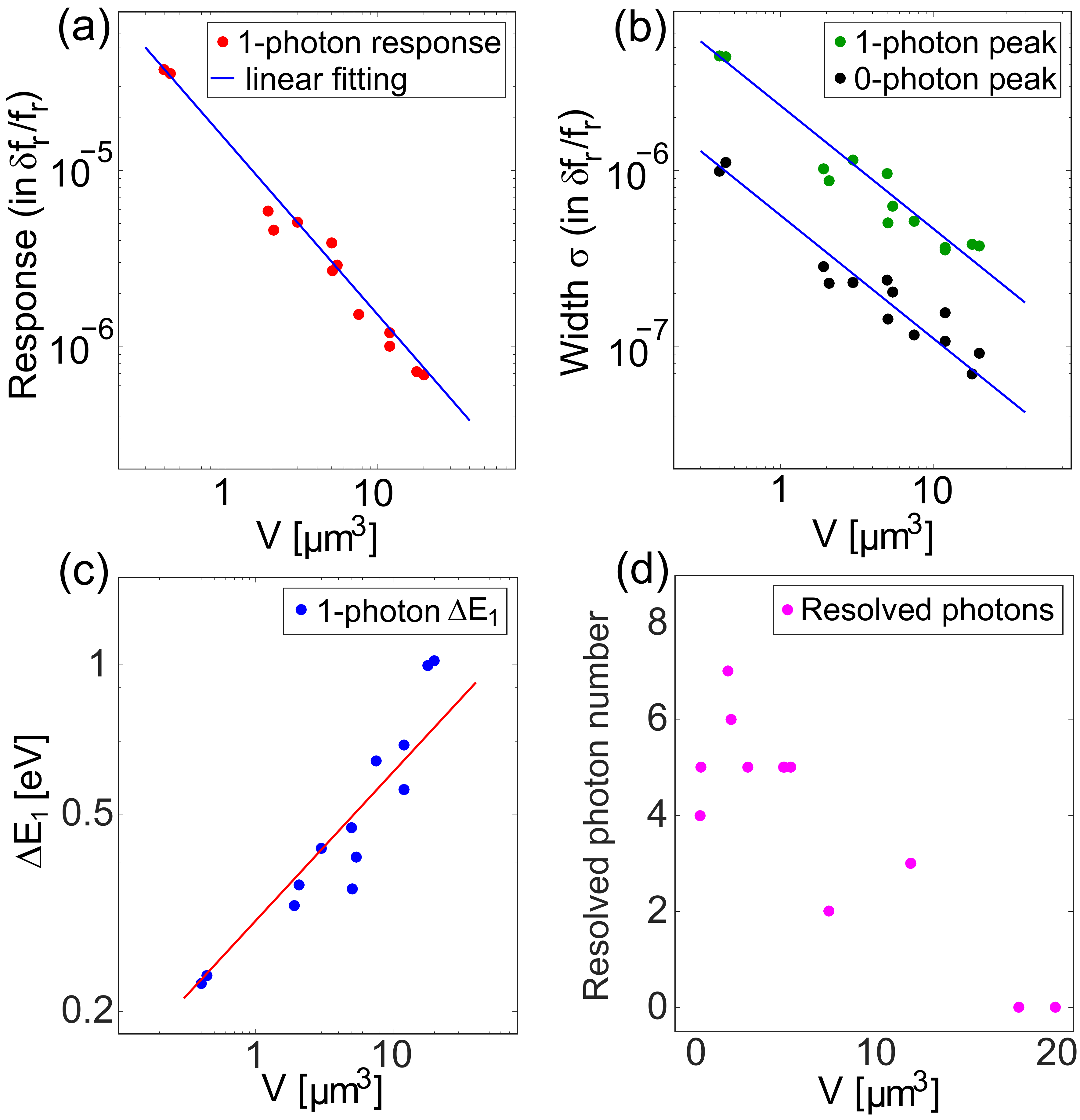}
\caption{[Color online] Log-log plot of 1-photon responsivity vs. absorber volume $V$. The data fit onto a straight line with a slope of -1 (blue line), indicating the measured responsivity is proportional to 1/$V$. (b) Log-log plot of the widths of 0-photon and 1-photon peak vs. $V$. The data points fall into two groups and both can be fitted by straight lines (the two blue lines) with the same slope of -0.7, suggesting both widths roughly scale as $V^{-0.7}$. (c) Log-log plot of 1-photon energy resolution $\Delta E_{1}$ vs. $V$ and the fitted red line indicates a $V^{0.3}$ scaling of $\Delta E_{1}$.
%The fluctuations in this trend indicate that the absorber volume is not the only factor to determine $\Delta E_{1}$.
(d) The maximum number of resolved photons $N_r$ vs. $V$.}
\label{fig:cavity}
\end{figure}

Fig.~3(b) shows the widths (i.e., the standard deviations $\sigma_{0}$ and $\sigma_{1}$ converted to $\delta f_{r}/f_{r}$ which is a measure of the frequency noise) of the 0-photon peak (black dots) and 1-photon peak (green dots) as a function of $V$. Both widths roughly fit onto a power-law of $V^{-0.7}$ and the 1-photon peak is about $\sim$ 4.5 times wider than the 0-photon peak. Combining the responsivity data from Fig.~3(a) and the noise data from Fig.~3(b), we derive the 1-photon energy resolution $\Delta E_{1}$ from Eqn. (1) as a function of $V$, which is plotted in Fig.~3(c). We see that $\Delta E_{1}$ increases with $V$ and scales as  $\approx V^{0.3}$. Our results suggest that the energy resolution improves as the absorber volume is reduced. The best $\Delta E_{1}$ we obtained is 0.22 eV, corresponding to an energy-resolving power of $R = h\nu/\Delta E_{1} = 3.7$ at 1550~nm, which is achieved in the resonator with the smallest absorber volume of 0.4~$\mu$m$^3$ and also the narrowest inductor width of 1~$\mu$m. In Fig.~3(d), we plot the maximum number of photons that can be resolved by each detector $N_r$ as a function of its absorber volume $V$. We see that $N_r$ drops at both smallest and largest $V$. $N_r$ drops at large $V$ because the energy resolution degrades as $V$ is increased (Fig.~3(c)). $N_r$ also drops at small $V$ because the large responsivity and high photon number lead to ``saturation'' of the detector, where the frequency shift of the pulse exceeds the resonator bandwidth and the signal-to-noise ratio is degraded.
%For example, for the resonator with inductor width of 1~$\mu$m and the smallest volume of 0.4~$\mu$m$^3$, the response pulse height (in phase angle) is about 0.93 radians per single-photon absorption. Thus when the absorbed photon number is more than 2, we are into the non-linear phase regime. Note that in our data analysis, we have used a rigorous non-linear fitting procedure to convert the phase shift to fractional frequency shift.
To increase the bandwidth for operation,  we can design resonator with lower $Q_{c}$ and/or higher resonance frequency $f_r$.

%For the 2 resonators with $V$ = 1.92~$\mu$m$^3$ and 2.08~$\mu$m$^3$ (both with absorber width of 2 $\mu$m), up to 7-photon event can be resolved. The exception is that for the 2 resonators with the smallest $V$ = 0.4~$\mu$m$^3$ and 0.44~$\mu$m$^3$ (both with absorber width 1 $\mu$m), the resolved photon number drops below 6 though they have the lowest energy resolution, which is possibly due to the finite operation range in the resonance loop. High-number photon event with high responsivity will easily put the resonator to the phase change insensitive region in the resonance loop, limiting the maximum number of photons which can be counted.

 %For the same mean photon number (but not same optical power due to different coupling efficient)
 %With increasing optical power, we can resolve up to 8-photon event. The reason not able to count higher number of photons are due to 1). noise and 2).finite operation range.

 %Both the energy resolution and resolved photon number are comparable to TES.

%High energy impacts...

The best theoretical energy-resolving power that can be achieved by a MKID as a pair-breaking detector is given by $R= \frac{1}{2.355}\sqrt{\frac{\eta h\nu }{F\Delta}}$, where $\eta \approx 0.57$ is the conversion efficiency from photons to quasiparticles~\cite{STJ}, $h\nu$ is the energy of the incident photons, $\Delta = 1.72 k_{B}T_{c}$  is the superconducting gap energy of the absorber material, and $F$ is the Fano factor~\cite{Fano}. This predicts a theoretical $R = 45$ at 1550~nm (a typical value of $F = 0.2$ is assumed), which is an order of magnitude higher than the $R = 3.7$ achieved by our best detector. Coincidentally, the optical lumped-element MKIDs~\cite{Marsden,Marsden2} made from 20-60~nm substoichiometric TiN films have a typical energy-resolving power $R = 16$ at 254~nm, which is also an order of magnitude below their Fano limit $R = 150$. While this suggests that TiN-based photon counting detectors have large room to improve, it is important to understand why they ``underperform'' their theoretical prediction. In fact, $\eta\approx0.57$ is the ideal conversion efficiency when photons are absorbed in a bulk superconductor. Our film is only $20$~nm thick and the high energy phonons may quickly escape the film into the substrate before breaking more quasiparticles, leading to a efficiency $\eta$ smaller than $0.57$ and a smaller response. This phonon loss process may also fluctuate and cause additional noise, as observed in the thin film superconducting tunnel junction photon detectors~\cite{STJ}. In future experiments, we plan to futher explore this phonon loss effect, as well as the $V^{0.3}$ energy resolution scaling, by testing different thickness of TiN films and by making the absorber on a suspended membrane.

Many aspects in our design and experimental setup can be improved. If the responsivity and noise trends still hold below 0.4~$\mu$m$^3$, we expect that better energy resolution can be achieved by using an absorber volume even smaller than 0.1 $\mu$m$^3$. Instead of using $T_{c} \approx$ 1.4~K trilayer, a lower $T_{c}$ TiN film with a lower gap energy may further boost the responsivity. Suspending the absorber on a thin silicon membrane may increase the quasiparticle recombination time and the conversion efficiency, as suggested by the ``phonon recycling" scheme \cite{Fyhrie,Ulbricht}. According to the optical measurement on thin TiN films by Volkonen\cite{Valkonen}, we estimate that the reflectance and transmittance for our 20~nm TiN film are about 60$\%$ and 10$\%$ respectively, indicating approximately only 30$\%$ photons are absorbed. The photon absorption efficiency can be greatly enhanced by adding anti-reflection coating and embedding the absorber in an optical structure~\cite{Lita2}. To efficiently collect every photon, the input light should be precisely confined onto the absorber active area, which can be realized using advanced alignment and coupling techniques, such as direct fiber coupling to the detector \cite{Miller2} or through a fusion-spliced microlens~\cite{Dauler2}.

%IDC can be coated with a reflection-layer to reduce the photon absorption into the non-absorber parts;

%Second, efforts can be made to achieve a more uniform photon response. For exmaple, IDC can be covered by a layer of Nb instead of Al, and then coated with a reflection-layer. This will help reducing the photon absorption by IDC and also lowering the unwanted response after photon absorbed by IDC.

%We will explore these and evaluate the detection efficiency of our device in the next experiment.

In conclusion, we have demonstrated photon counting at 1550~nm using TiN/Ti/TiN trilayer MKIDs.
%and achieved performance comparable to TES.
Energy resolution as low as $\Delta E \approx$ 0.22~eV is obtained and up to 7-photon events can be resolved. By studying devices with a variety of geometries, we have systematically investigated the dependence of photon counting performance on the absorber volume. The energy resolution improves as the absorber volume is reduced.
%find that the device responsivity is inversely proportional to the absorber volume, resulting that both the energy-resolving power and the maximum countable photon number decrease with increasing absorber volume.
Further improvements in these detectors are possible by improving the detector design and optimizing the optical coupling to maximize the photon absorption into the absorber. With the energy resolution of our MKID photon counting detectors approaching the performance of TESs (currently a factor of two better), the multiplexing advantage of MKIDs may stand out in applications where a large array of detectors with high photon-resolving power is needed.

%In conclusion, we have demonstrated photon counting in TiN MKIDs with performance comparable to TES. Experimental results show that our device can resolve up to 7-photon events and achieve an energy resolution of 0.22 eV. In this paper, we focus on the relationship of the absorber volume $V$ to the energy-resolving power and the resolved photon number. Exactly, we find that both the energy-resolving power and the resolved photon number decrease with increasing $V$. But if the absorber volume $V$ is too small, the high-number photon event will easily make the response to the insensitive region of the resonance loop, limiting the maximum number of resolved photons. On the other hand, we have not made attempt to improve the optical coupling or quantum efficiency in current device, which requires further research. As MKIDs develop, they may find more extensive application.

\vspace{1em}

The MKID devices were fabricated in the NIST-Boulder microfabrication facility. This work was supported in part by the National Natural Science Foundation of China (Grant Nos. 61301031, U1330201). L. F. Wei thanks Profs. C. D. Xie and K. C. Peng for their encouragements and useful discussions.

%%%%%%%%%%%%%%%%%%%%%%%%%%%%%%%%%%%%%%%%%%%%%%%%%%%%%%%%%%%%%

\newpage


\begin{thebibliography}{1}
\bibitem{Hiskett} P. Hiskett, D. Rosenberg, C. Peterson, R. Hughes, S. Nam, A. Lita, A. Miller, and J. Nordholt, New J. Phys. \textbf{8}, 193 (2006).
\bibitem{Knill} E. Knill, R. Laflamme, and G. J. Milburn, Nature \textbf{409}, 46 (2012).
\bibitem{Zwinkels} J. Zwinkels, E. Ikonen, N. Fox, G. Ulm, and M. Rastello, Metrologia \textbf{47}, R15 (2010).
%\bibitem{Pomarico} E. Pomarico, B. Sanguinetti, R. Thew, and H. Zbinden, Optics Express \textbf{18}, 10750 (2010).
\bibitem{Finger} G. Finger, I. Baker, D. Alvarez, D. Ives, L. Mehrgan, M. Meyer, J. Stegmeier and H. J. Weller, Proc. of SPIE \textbf{9148}, 914817 (2014).
\bibitem{Tsman} G. Gol'tsman, O. Okunev, G. Chulkova, A. Lipatov, A. Semenov, K. Smirnov, B. Voronov, A. Dzardanov, C. Williams and R. Sobolewski, Appl. Phys. Lett. \textbf{79}, 705 (2001).
\bibitem{Divochiy} A. Divochiy, F. Marsili, D. Bitauld, A. Gaggero, R. Leoni, F. Mattioli, A. Korneev, V. Seleznev, N. Kaurova, O. Minaeva, G. Gol'tsman, K. Lagoudakis, M. Benkhaoul, F. Levy, and A. Fiore, Nature Photonics \textbf{2}, 302 (2008).
\bibitem{Dauler} E. Dauler, A. Kerman, B. Robinson, J. Yang, B. Voronov, G. Goltsman, S. Hamilton, and K. Berggren, Journal of Modern Optics \textbf{56}, 364 (2009).
\bibitem{Mattioli} F. Mattioli, Z. Zhou, A. Gaggero, R. Gaudio, R. Leoni, and A. Fiore, Optics Express \textbf{24}, 9067 (2016).
\bibitem{Irwin} K. D. Irwin, Appl. Phys. Lett. \textbf{66}, 1998 (1995).
\bibitem{Miller} A. J. Miller, S. W. Nam, J. M. Martinis, and A. V. Sergienko, Appl. Phys. Lett. \textbf{83}, 791 (2003).
\bibitem{Lita} A. Lita, A. Miller and S. Nam, Optics Express \textbf{16}, 3032 (2008).
\bibitem{Calkin} B. Calkins, P. Mennea, A. Lita, B. Metcalf, W. Kolthammer, A. Linares, J. Spring, P. Humphreys, R. Mirin, J. Gates, \emph{et al.}, Optics Express \textbf{21}, 22657 (2013).
\bibitem{Lolli} L. Lolli, E. Taralli, and M. Rajteri, J Low Temp Phys \textbf{167}, 803 (2012).
\bibitem{Lolli2} G. Brida, L. Ciavarella, I. Degiovanni, M. Genovese, L. Lolli, M. Mingolla, F. Piacentini, M. Rajteri, E. Taralli, and M. Paris, New. J. Phys. \textbf{14}, 085001 (2012).
\bibitem{Lolli3} L. Lolli, E. Taralli, C. Portesi, E. Monticone, and M. Rajteri, Appl. Phys. Lett. \textbf{103}, 041107 (2013).
\bibitem{Day}P. K. Day, H. G. LeDuc, B. A. Mazin, A. Vayonakis, and J. Zmuidzinas, Nature \textbf{425}, 817 (2003).
\bibitem{zmuidzinas} J. Zmuidzinas, Annual Review of Condensed Matter Physics \textbf{3}, 169 (2012).
\bibitem{yiwen} Y. Wang, P. Zhou, L. Wei, H. Li, B. Zhang, M. Zhang, Q. Wei, Y. Fang, and Chunhai Cao, J. Appl. Phys. \textbf{114}, 153109 (2013).
\bibitem{Bumble} B. A. Mazin, B. Bumble, S. R. Meeker, K. O¡¯Brien, S. McHugh, and E. Langman, Opt. Express \textbf{20}, 1503 (2012).
\bibitem{Jiansong} J. Gao, M. Vissers, M. Sandberg, F. Silva, S. Nam, D. Pappas, D. Wisbey, E. Langman, S. Meeker, B. Mazin  H. Leduc, J. Zmuidzinas, and K. Irwin, Appl. Phys. Lett. \textbf{101}, 142602 (2012).
\bibitem{Vissers} M.R. Vissers, J. Gao, M. Sandberg, S. M. Duff, D. S. Wisbey, K.D. Irwin, and D. P. Pappas. Appl. Phys. Lett. \textbf{102}, 232603 (2013).
\bibitem{hannes}J. Hubmayr, J. Beall, D. Becker, H.-M. Cho, M. Devlin, B. Dober, C. Groppi, G. C. Hilton, K. D. Irwin, D. Li, P. Mauskopf, D. P. Pappas, J. Van Lanen, M. Vissers, Y. Wang, L. F. Wei, and J. Gao, Appl. Phys. Lett. \textbf{106}, 073505 (2015).

\bibitem{gao2008}J. Gao, M. Daal, J. Martinis, A. Vayonakis, J. Zmuidzinas, B. Sadoulet, B. Mazin, P. Day, and H. Leduc, Appl. Phys. Lett. \textbf{92}, 212504 (2008).

\bibitem{PropTiN} J. Gao, M. R. Vissers, M. Sandberg, D. Li, H. M. Cho, C. Bockstiegel, B. A. Mazin, H. G. Leduc, S. Chaudhuri, D. P. Pappas,and K. D. Irwin, J Low Temp Phys, \textbf{176}, 136 (2014).
\bibitem{gaothesis} J. Gao, Ph.D. thesis, Caltech, 2008.
\bibitem{STJ} D. D. E. Martin, P. Verhoeve, A. Peacock, A. G. Kozorezov, J. K. Wigmore, H. Rogalla, and R. Venn, Appl. Phys. Lett. \textbf{88}, 123510 (2006).

\bibitem{Fano}U. Fano,  Phys. Rev. \textbf{72}, 26(1947).
\bibitem{Marsden}D. Marsden, B. A. Mazin, B. Bumble, Seth Meeker, K. O'Brien, S. McHugh, M. Strader, and E. Langman, Proc. SPIE  \textbf{8453}, 84530B (2012).
\bibitem{Marsden2}B. A. Mazin, S. R. Meeker, M. J. Strader, P. Szypryt, D. Marsden, J. C. van Eyken, G. E. Duggan, A. B. Walter, G. Ulbricht, M. Johnson, B. Bumble, K. O'Brien, and C. Stoughton, Publications of the Astronomical Society of the Pacific, \textbf{125}, 1348 (2013).
\bibitem{Fyhrie}A. Fyhrie, C. McKenney, J. Glenn, H. G. LeDuc, J. Gao, P. Day, and Jonas Zmuidzinas, Proc. SPIE  \textbf{9914}, 99142B (2012).
\bibitem{Ulbricht}G. Ulbricht, B. A. Mazin, P. Szypryt, A. B. Walter, C. Bockstiegel, and B. Bumble, Appl. Phys. Lett. \textbf{106}, 251103 (2015).
\bibitem{Valkonen} E. Valkonen, C. Ribbing, and J. Sundgren, Applied Optics \textbf{25}, 3624 (1986).
\bibitem{Lita2} A. Lita, B. Calkins, L. Pellouchoud, A. Miller, and S. Nam, Proc. of SPIE \textbf{7681}, 76810D-1 (2010).
\bibitem{Miller2} A. Miller, A. Lita, B. Calkins, I. Vayshenker, S. Gruber, and S. Nam, Optics Express \textbf{19}, 9102 (2011).
\bibitem{Dauler2} E. A. Dauler, M. E. Grein, A. J. Kerman, F. Marsili, S. Miki, S. W. Nam, M. D. Shaw, H. Terai, V. B. Verma, and T. Yamashita, Opt. Eng.  \textbf{53}, 081907 (2014).


\end{thebibliography}
\end{document}